\documentclass[apj,numberedappendix,twocolappendix,appendixfloats]{emulateapj}
\usepackage{apjfonts}
\usepackage[breaklinks,colorlinks,citecolor=blue,linkcolor=magenta]{hyperref} 
\newcommand{\Ha}{$\rm{H} \alpha$}

\newcommand{\etal}{et~al.}
\newcommand{\PVdblt}{{\rm P}\kern 0.1em{\sc v}~$\lambda\lambda 1117, 1128$}
\newcommand{\CaIIdblt}{{\rm Ca}\kern 0.1em{\sc ii}~$\lambda\lambda 3934, 3969$}
\newcommand{\AlIIIdblt}{{\rm Al}\kern 0.1em{\sc iv}~$\lambda\lambda 1855, 1863$}
\newcommand{\CIVdblt}{{\rm C}\kern 0.1em{\sc iv}~$\lambda\lambda 1548, 1550$}
\newcommand{\MgIIdblt}{{\rm Mg}\kern 0.1em{\sc ii}~$\lambda\lambda 2796, 2803$}
\newcommand{\NVdblt}{{\rm N}\kern 0.1em{\sc v}~$\lambda\lambda 1238, 1242$}  
\newcommand{\SVIdblt}{{\rm S}\kern 0.1em{\sc vi}~$\lambda\lambda 933, 944$} 
\newcommand{\OVIdblt}{{\rm O}\kern 0.1em{\sc vi}~$\lambda\lambda 1031, 1037$} 
\newcommand{\SiIIdblt}{{\rm Si}\kern 0.1em{\sc ii}~$\lambda\lambda 1190, 1193$} 
\newcommand{\SiIVdblt}{{\rm Si}\kern 0.1em{\sc iv}~$\lambda\lambda 1393, 1402$} 
\newcommand{\PV}{\hbox{{\rm P}\kern 0.1em{\sc v}}}
\newcommand{\AlI}{\hbox{{\rm Al}\kern 0.1em{\sc i}}}
\newcommand{\AlII}{\hbox{{\rm Al}\kern 0.1em{\sc ii}}}
\newcommand{\AlIII}{{\hbox{\rm Al}\kern 0.1em{\sc iii}}}
\newcommand{\CaII}{\hbox{{\rm Ca}\kern 0.1em{\sc ii}}}
\newcommand{\CII}{\hbox{{\rm C}\kern 0.1em{\sc ii}}}
\newcommand{\CIIe}{\hbox{{\rm C$^{\ast}$}\kern 0.1em{\sc ii}}}
\newcommand{\CIII}{\hbox{{\rm C}\kern 0.1em{\sc iii}}}
\newcommand{\CIV}{\hbox{{\rm C}\kern 0.1em{\sc iv}}}
\newcommand{\CV}{\hbox{{\rm C}\kern 0.1em{\sc v}}}
\newcommand{\HI}{\hbox{{\rm H}\kern 0.1em{\sc i}}}
\newcommand{\HII}{\hbox{{\rm H}\kern 0.1em{\sc ii}}}
\newcommand{\Lya}{\hbox{{\rm Ly}\kern 0.1em$\alpha$}}
\newcommand{\Lyb}{\hbox{{\rm Ly}\kern 0.1em$\beta$}}
\newcommand{\Lyg}{\hbox{{\rm Ly}\kern 0.1em$\gamma$}}
\newcommand{\Lyd}{\hbox{{\rm Ly}\kern 0.1em$\delta$}}
\newcommand{\Lye}{\hbox{{\rm Ly}\kern 0.1em$\epsilon$}}
\newcommand{\Lyphi}{\hbox{{\rm Ly}\kern 0.1em$\phi$}}
\newcommand{\Lyfive}{\hbox{{\rm Ly}\kern 0.1em$5$}}
\newcommand{\Lysix}{\hbox{{\rm Ly}\kern 0.1em$6$}}
\newcommand{\Lyseven}{\hbox{{\rm Ly}\kern 0.1em$7$}}
\newcommand{\Lyeight}{\hbox{{\rm Ly}\kern 0.1em$8$}}
\newcommand{\Lynine}{\hbox{{\rm Ly}\kern 0.1em$9$}}
\newcommand{\Lyten}{\hbox{{\rm Ly}\kern 0.1em$10$}}
\newcommand{\Lyeleven}{\hbox{{\rm Ly}\kern 0.1em$11$}}
\newcommand{\HeI}{\hbox{{\rm He}\kern 0.1em{\sc i}}}
\newcommand{\HeII}{\hbox{{\rm He}\kern 0.1em{\sc ii}}}
\newcommand{\FeI}{\hbox{{\rm Fe}\kern 0.1em{\sc i}}}
\newcommand{\FeII}{\hbox{{\rm Fe}\kern 0.1em{\sc ii}}}
\newcommand{\FeIII}{\hbox{{\rm Fe}\kern 0.1em{\sc iii}}}
\newcommand{\MnII}{\hbox{{\rm Mn}\kern 0.1em{\sc ii}}}
\newcommand{\MgI}{\hbox{{\rm Mg}\kern 0.1em{\sc i}}}
\newcommand{\MgIb}{\hbox{{\rm Mg}\kern 0.1em{\sc i}}\kern 0.05em{\rm b}}
\newcommand{\MgII}{\hbox{{\rm Mg}\kern 0.1em{\sc ii}}}
\newcommand{\MgIII}{\hbox{{\rm Mg}\kern 0.1em{\sc iii}}}
\newcommand{\NI}{\hbox{{\rm N}\kern 0.1em{\sc i}}}
\newcommand{\NII}{\hbox{{\rm N}\kern 0.1em{\sc ii}}}
\newcommand{\NIII}{\hbox{{\rm N}\kern 0.1em{\sc iii}}}
\newcommand{\NV}{\hbox{{\rm N}\kern 0.1em{\sc v}}}
\newcommand{\OVI}{\hbox{{\rm O}\kern 0.1em{\sc vi}}}
\newcommand{\OI}{\hbox{{\rm O}\kern 0.1em{\sc i}}}
\newcommand{\OII}{\hbox{[{\rm O}\kern 0.1em{\sc ii}]}}
\newcommand{\OIII}{\hbox{[{\rm O}\kern 0.1em{\sc iii}]}}
\newcommand{\OIV}{\hbox{{\rm O}\kern 0.1em{\sc iv}]}}
\newcommand{\SI}{{\rm S}\kern 0.1em{\sc i}}
\newcommand{\SIV}{{\rm S}\kern 0.1em{\sc iv}}
\newcommand{\SVI}{{\rm S}\kern 0.1em{\sc vi}}
\newcommand{\SiI}{\hbox{{\rm Si}\kern 0.1em{\sc i}}}
\newcommand{\SiII}{\hbox{{\rm Si}\kern 0.1em{\sc ii}}}
\newcommand{\SiIII}{\hbox{{\rm Si}\kern 0.1em{\sc iii}}}
\newcommand{\SiIV}{\hbox{{\rm Si}\kern 0.1em{\sc iv}}}
\newcommand{\SII}{\hbox{{\rm S}\kern 0.1em{\sc ii}}}
\newcommand{\SIII}{\hbox{{\rm S}\kern 0.1em{\sc iii}}}
\newcommand{\NaI}{\hbox{{\rm Na}\kern 0.1em{\sc i}}}
\newcommand{\NaID}{\hbox{{\rm Na}\kern 0.1em{\sc i}}\kern 0.05em{\rm D}}
\newcommand{\TiII}{\hbox{{\rm Ti}\kern 0.1em{\sc ii}}}
\newcommand{\kms}{\hbox{~km~s$^{-1}$}}

\slugcomment{} 
\shorttitle{\sc ISM \& CGM Metallicities}
\shortauthors{\sc Kacprzak et~al.}

\begin{document}


\title{The Relationship Between Galaxy ISM and Circumgalactic Gas
  Metallicities}


\author{\sc Glenn G. Kacprzak\altaffilmark{1}, Stephanie
  K. Pointon\altaffilmark{1,2}, Nikole M. Nielsen\altaffilmark{1},
  Christopher W. Churchill\altaffilmark{3}, Sowgat
  Muzahid\altaffilmark{2}, Jane C. Charlton\altaffilmark{4}}
                                                                   
\altaffiltext{1}{Swinburne University of Technology, Victoria 3122, Australia {\tt gkacprzak@swin.edu.au}}
\altaffiltext{2}{ARC Centre of Excellence for All Sky Astrophysics in 3 Dimensions (ASTRO 3D)}
\altaffiltext{3}{Leiden Observatory, Leiden University, P.O. Box 9513, 2300 RA Leiden, The Netherlands}
\altaffiltext{4}{New Mexico State University, Las Cruces, NM 88003, USA}
\altaffiltext{5}{The Pennsylvania State University, State College, PA 16801, USA}

\begin{abstract}
  We present ISM and CGM metallicities for 25 absorption systems
  associated with isolated star-forming galaxies
  ($\left<z\right>=0.28$) with
  9.4$\leq$log(M$_*$/M$_{\odot}$)$\leq$10.9 and with absorption
  detected within 200~kpc. Galaxy ISM metallicities were measured
  using {\Ha}/[\NII] emission lines from Keck/ESI spectra. CGM
  single-phase low-ionization metallicities were modeled using MCMC
  and Cloudy analysis of absorption from {\it HST}/COS and Keck/HIRES
  or VLT/UVES quasar spectra. We find that the star-forming galaxy ISM
  metallicities follow the observed stellar mass metallicity relation
  ($1\sigma$ scatter 0.19~dex). CGM metallicity shows no dependence
  with stellar mass and exhibits a scatter of $\sim$2~dex. All CGM
  metallicities are lower than the galaxy ISM metallicities and are
  offset by log($dZ)=-1.17\pm0.11$.  There is no obvious metallicity
  gradient as a function of impact parameter or virial radius
  ($<2.3\sigma$ significance). There is no relationship between the
  relative CGM--galaxy metallicity and azimuthal angle. We find the
  mean metallicity differences along the major and minor axes are
  $-$1.13$\pm$0.18 and $-$1.23$\pm$0.11, respectively. Regardless of
  whether we examine our sample by low/high inclination or low/high
  impact parameter, or low/high N({\HI}), we do not find any
  significant relationship with relative CGM--galaxy metallicity and
  azimuthal angle.  We find that 10/15 low column density systems
  (logN({\HI})$<17.2$) reside along the galaxy major axis while high
  column density systems (logN({\HI})$\geq17.2$) reside along the
  minor axis. This suggest N({\HI}) could be a useful indicator of
  accretion/outflows. We conclude that CGM is not well mixed, given
  the range of galaxy-CGM metallicities, and that metallicity at low
  redshift might not be a good tracer of CGM processes. On
  the-other-hand, we should replace integrated line-of-sight, single
  phase, metallicities with multi-phase, cloud-cloud metallicities,
  which could be more indicative of the physical processes within the
  CGM.
\end{abstract}




\section{Introduction}
\label{sec:intro}

It is undeniable that there exists a massive reservoir of multi-phase
gas that resides around star-forming galaxies \citep{tumlinson17}.  A
large collection of works provide evidence that outflows and accretion
are ongoing processes that continuously change the properties of the
circumgalactic medium (CGM) and the host galaxies.  Low ionization
level ions within the CGM show strong kinematic signatures that are
consistent with large-scale outflows
\citep{bouche06,tremonti07,martin09,weiner09,
  nestor11,noterdaeme10,coil11,kacprzak10,kacprzak14,rubin10,menard12,
  martin12,noterdaeme12,krogager13,peroux13,rubin14,crighton15,
  nielsen15,nielsen16,lan18}. Furthermore, low angular momentum and
co-rotating gas around galaxies and orientation dependant absorption
velocity widths point to signatures of gas accretion \citep{steidel02,
  kacprzak10,ho17,kacprzak17,martin19,zabl19}.

The CGM, as traced by {\MgII} absorption, appears to have a preference
to exist along the major and minor axes of galaxies
\citep{bouche12,kacprzak12a,schroetter19}, while the equivalent width
of the absorption is highest along the galaxy minor axis
\citep{bordoloi11,kacprzak12a,lan14,lan18}. This geometric dependence
could be additional evidence for outflows and accretion. Furthermore,
the metallicity distribution bimodality found for $z<0.4$ Lyman limit systems
(LLS) and partial Lyman limit systems (pLLS) shows a high
([X/H]$\sim-0.4$) and a low ([X/H]$\sim-1.7$) metallicity peak that
could be attributed to being caused by outflows and accretion
\citep{lehner13,lehner19,wotta16,wotta19}.  Thus the spatial
distribution of metallicity around galaxies would seem to be a likely
key to understanding the origins of the CGM.

 Numerous studies have obtained the CGM metallicity associated with a
  known galaxy in an effort to determine the origin and history of the
  absorption.  These CGM metallicities generally reflect the
  metallicity bimodality where systems near galaxies are either
  metal-poor with metallicities between $-2<$[X/H]$<-1$
  \citep{tripp05,cooksey08,kacprzak10b,ribaudo11,thom11,churchill12,bouche13,crighton13,stocke13,kacprzak14,crighton15,muzahid15,bouche16,fumagalli16b,peroux16}
  or metal-enriched with metallicities of [X/H]$>-0.5$
  \citep{chen05,peroux11,krogager13,stocke13,crighton15,muzahid15,muzahid16,peroux16}. However,
  little is known about the host galaxy geometry with respect to the
  quasar sight-line in most cases.  Using a sample of 47 galaxies with
  measured morphologies/orientations and CGM metallicities,
  \citet{pointon19} has shown that CGM metallicities do not
  correlate with azimuthal angle or inclination of the galaxy
  regardless of impact parameter and N({\HI}). Thus, it is possible
  that the spatial azimuthal dependence and the metallicity bimodality
  are unrelated.  

  However, we do not fully understand how the host galaxy-ISM
  metallicities relate to the CGM metallicities. Since CGM galaxies
  span a range of stellar mass and given that there is a well-known
  galaxy stellar mass and ISM metallicity relation found at all
  redshifts
  \citep[e.g,][]{tremonti04,sanders14,steidel14,zahid14,kacprzak15b},
  then it is possible that the difference between the ISM and CGM
  metallicities would be more telling of the origins of the
  CGM. 

  Initial work from \citet{prochaska17} has shown that for $\sim20$
  systems, the CGM metallicity does not correlate with the ISM
  metallicity of host galaxies. In addition, work by \citet{peroux16}
  examined the metallicity difference between the galaxy ISM
  metallicity and CGM metallicity for nine systems.  They found at low
  azimuthal angles, there are a range of ISM-CGM metallicity
  differences which would be unexpected for accreting gas. They only
  had two lower ISM-CGM metallicity systems along the minor axis,
  which is also unexpected for an outflow model.  Fully exploring the
  relative galaxy ISM-CGM metallicities could provide additional
  insight into the relationship between galaxies and ongoing processes
  within the CGM.

  We aim to further explore the relationship between the galaxy ISM
  and CGM metallicities and how they relate to the expectations of
  accretion/outflow models. We have acquired Keck/ESI spectra for 25
  star-forming galaxies to obtain their ISM metallicities and their
  CGM metallicities are derived in \citet{pointon19}. We examine the
  stellar mass-metallicity relation for the galaxies and the CGM and
  test if the relative metallicity difference, defined to be the
  difference between the ISM and CGM metallicities, is dependent on
  hydrogen column density and/or galaxy properties such as azimuthal
  angle, inclination angle, and impact parameter.  In
  Section~\ref{sec:data} we present our sample, data and data
  reduction. In Section~\ref{sec:results} we present our observational
  results. In Section~\ref{sec:discussion}, we discuss what can be
  inferred from the results and concluding remarks are offered in
  Section~\ref{sec:conclusion}.  Throughout we adopt an H$_{\rm
    0}=70$~\kms Mpc$^{-1}$, $\Omega_{\rm M}=0.3$,
  $\Omega_{\Lambda}=0.7$ cosmology.

%%%%%%%%%%%%%%%%%%%%%%%%%%%%%%%%%%%%%%%%%%%%%%%%%%%%%%%%%%%%%%%%%%%%%%%%%%%%%
\begin{deluxetable*}{llcrcrclrrrrr}
\tabletypesize{\scriptsize}
\tablecaption{Absorption and host galaxy properties\label{tab:morph}}
\tablecolumns{13}
\tablewidth{0pt} 
\tablehead{
\colhead{Quasar\tablenotemark{ a}}&
\colhead{$z_{\rm gal}$\tablenotemark{ b}} &
\colhead{M$_{r}$} &
\colhead{R$_{vir}$} &
\colhead{log(M$_{h}$)} &
\colhead{log(M$_{*}$)} &
\colhead{log(O$/$H)$+$12\tablenotemark{b}} &
\colhead{$i$} &
\colhead{$\Phi$} &
\colhead{$D$} &
\colhead{log N({\HI})} &
\colhead{log N({\HI})} &
\colhead{log($Z_{CGM}$)}\\
\colhead{field }&
\colhead{ } &
\colhead{(AB)} &
\colhead{(kpc)} &
\colhead{(M$_{\odot}$)} &
\colhead{(M$_{\odot}$)} &
\colhead{} &
\colhead{(degree) } & 
\colhead{(degree) } & 
\colhead{(kpc) } & 
\colhead{$_{Measured}$ (cm$^{-2}$) } &
\colhead{$_{Modeled}$ (cm$^{-2}$) } &
\colhead{[Si/H] ($Z_{\odot}$) }
}
\startdata
J012528 &0.398525  & $-$21.99 &  $285.5_{-32}^{+37}$&$12.5_{-0.2}^{+0.2}$& $10.9_{-0.2}^{+0.2}$ & 8.69  & $63.2_{-2.6}^{+1.7}$   & $73.4_{-4.7}^{+4.6}$ & $163.0$&$  [18.85,19.00]        $ & $18.85^{+0.04}_{-0.01}$ &$ 	-1.56	^{+0.03	}_{-0.03} $  \\[+0.35ex]   
J035128 &0.356992  & $-$20.86 &  $190.9_{-26}^{+48}$&$12.0_{-0.2}^{+0.3}$& $10.4_{-0.2}^{+0.3}$ & 8.63  & $28.5_{-12.5}^{+19.8}$ & $4.9_{-40.2}^{+33.0}$ &$72.3$  &$  16.86	\pm0.03 $ &$16.86^{+0.03}_{-0.03}$ & $ 	-0.38	^{+0.04	}_{-0.04} $ \\[+0.35ex]   
J040748 &0.495164  & $-$19.73 &$124.4_{-18}^{+52}$&$11.4_{-0.2}^{+0.5}$& $9.7_{-0.2}^{+0.5}$    &8.46 &$67.2_{-7.5}^{+7.6}$      & $21.0_{-3.7}^{+5.3}$ & 107.6   &$  14.34	\pm0.56 $ &$14.35^{+0.35}_{-0.35}$ & $ 	-1.10	^{+0.49	}_{-0.55} $ \\[+0.35ex] % ESIJan2016  8.4621
J045608 &0.277938  & $-$19.12 &  $122.0_{-18}^{+57}$&$11.4_{-0.2}^{+0.5}$& $ 9.8_{-0.2}^{+0.5}$ &8.14   & $71.2_{-2.6}^{+2.6}$   & $78.4_{-2.1}^{+2.1}$ & 50.7   &$  [15.06,19.00]        $ &$15.71^{+1.55}_{-0.73}$ & $ <	-1.40	   	 $ \\[+0.35ex]  % ESIDEC2014  8.135
J045608 &0.381511  & $-$20.87 &  $192.3_{-26}^{+48}$&$12.0_{-0.2}^{+0.3}$& $10.3_{-0.2}^{+0.3}$ & 8.67  & $57.1_{-2.4}^{+19.9}$  & $63.8_{-2.7}^{+4.3}$ & $103.4$&$  15.10	\pm0.39 $ &$15.13^{+0.38}_{-0.35}$ & $ 	-0.06	^{+0.03	}_{-1.01} $ \\[+0.35ex]   
J045608 &0.48382   & $-$21.91 &  $241.8_{-27}^{+38}$&$12.3_{-0.2}^{+0.2}$& $10.6_{-0.2}^{+0.2}$ &8.67   & $42.1_{-3.1}^{+3.1}$   & $85.2_{-3.7}^{+3.7}$ & 108.0  &$  [16.53,19.00]        $ &$17.65^{+0.18}_{-0.17}$ & $ 	-1.32	^{+0.15	}_{-0.15} $ \\[+0.35ex]  % K10ESI     8.6662
J085334 &0.163403  & $-$20.56 &  $167.6_{-24}^{+48}$&$11.9_{-0.2}^{+0.3}$& $10.3_{-0.2}^{+0.3}$ &8.86   & $70.1_{-0.8}^{+1.4}$   & $56.0_{-0.8}^{+0.8}$ & 26.2   &$  19.93	\pm0.01 $ &$19.93^{+0.01}_{-0.01}$ & $ 	-1.70	^{+0.06	}_{-0.05} $ \\[+0.35ex]  % ESIApr2014 8.8612
J091440 &0.244312  & $-$20.55 &  $170.7_{-24}^{+49}$&$11.9_{-0.2}^{+0.3}$& $10.3_{-0.2}^{+0.3}$ & 8.52  & $39.0_{-0.2}^{+0.4}$   & $18.2_{-1.0}^{+1.1}$ &   105.9&$  15.55	\pm0.03 $ &$15.55^{+0.04}_{-0.03}$ & $ 	-0.78	^{+0.09	}_{-0.10} $   \\[+0.35ex]   
J094331 &0.2284$^b$& $-$21.34 &  $216.5_{-27}^{+42}$&$12.2_{-0.2}^{+0.2}$& $10.6_{-0.2}^{+0.2}$ &8.94$^c$& $52.3_{-0.3}^{+0.3}$   & $30.4_{-0.4}^{+0.3}$ & 123.3 &$  16.03	\pm0.67 $ &$16.04^{+0.66}_{-0.48}$ & $ 	-1.33	^{+0.66	}_{-0.71} $  \\[+0.35ex]  % 8.94PPWerk  8.94 
J094331 &0.353052  & $-$19.88 &  $146.8_{-22}^{+54}$&$11.7_{-0.2}^{+0.4}$& $10.0_{-0.2}^{+0.4}$ & 8.53 & $44.4_{-1.2}^{+1.1}$    & $8.2_{-5.0}^{+3.0}$  &  $96.5$&$  16.46	\pm0.03 $ &$16.38^{+0.11}_{-0.01}$ & $ <	-1.69	   	 $  \\[+0.35ex]   
J095000 &0.211866  & $-$21.73 &  $246.9_{-29}^{+36}$&$12.4_{-0.2}^{+0.2}$& $10.8_{-0.2}^{+0.2}$ & 8.19 & $47.7_{-0.1}^{+0.1}$    & $16.6_{-0.1}^{+0.1}$ &  $93.6$&$  [16.28,19.00]        $ &$19.00^{+0.01}_{-0.09}$ & $ 	-1.48	^{+0.04	}_{-0.02} $   \\[+0.35ex]   
J100902 &0.227855  & $-$20.19 &  $154.5_{-23}^{+51}$&$11.8_{-0.2}^{+0.4}$& $10.1_{-0.2}^{+0.4}$ & 8.52 & $66.3_{-0.9}^{+0.6}$    & $89.6_{-1.3}^{+1.3}$ & $64.0$ &$  [17.51,19.00]        $ &$18.26^{+0.10}_{-0.13}$ & $ 	-2.00	^{+0.07	}_{-0.04} $  \\[+0.35ex]   
J113327 &0.154599  & $-$19.84 &  $138.8_{-21}^{+52}$&$11.6_{-0.2}^{+0.4}$& $10.0_{-0.2}^{+0.4}$ &8.19  & $23.5_{-0.2}^{+0.4}$    & $56.1_{-1.3}^{+1.7}$ & 55.6   &$  [15.82,17.00]        $ &$16.11^{+0.42}_{-0.29}$ & $ <	-1.98	   	 $ 	 \\[+0.35ex]    %ESIJan2016  8.192
J113910 &0.204194  & $-$19.99 &  $146.1_{-22}^{+52}$&$11.7_{-0.2}^{+0.4}$& $10.1_{-0.2}^{+0.4}$ & 8.67 & $81.6_{-0.5}^{+0.4}$    & $5.8_{-0.5}^{+0.4}$  &  $93.2$&$  [16.04,17.00]        $ &$16.04^{+0.04}_{-0.01}$ & $ 	-0.35	^{+0.03	}_{-0.07} $   \\[+0.35ex]   
J113910 &0.219724  & $-$17.67 &  $88.7 _{-14}^{+52}$&$11.0_{-0.2}^{+0.6}$& $ 9.4_{-0.2}^{+0.6}$ &8.37  & $85.0_{-8.5}^{+5.0}$    & $44.9_{-8.1}^{+8.9}$ & 122.0  &$  14.20	\pm0.07 $ &$14.30^{+0.01}_{-0.28}$ & $ <	0.63	   	 $	 \\[+0.35ex]    %ESIJan2016  8.3707
J113910 &0.319255  & $-$20.48 &  $170.4_{-24}^{+51}$&$11.9_{-0.2}^{+0.3}$& $10.2_{-0.2}^{+0.3}$ & 8.61 & $83.4_{-1.1}^{+1.4}$    & $39.1_{-1.7}^{+1.9}$ & $73.3$ & $  16.19	\pm0.03 $ & $16.19^{+0.03}_{-0.03}$ &$ 	-2.59	^{+0.58	}_{-0.04} $ \\[+0.35ex]   
J123304 &0.318757  & $-$20.62 &  $176.6_{-25}^{+50}$&$11.9_{-0.2}^{+0.3}$& $10.3_{-0.2}^{+0.3}$ &8.57  & $38.7_{-1.8}^{+1.6}$    & $17.0_{-2.3}^{+2.0}$ & 88.9   &$  15.72	\pm0.02 $ &$15.72^{+0.02}_{-0.02}$ & $ 	-1.14	^{+0.13	}_{-0.09} $ \\[+0.35ex]  % ESIJune2016 8.568
J124154 &0.205267  & $-$19.83 &  $140.2_{-21}^{+52}$&$11.6_{-0.2}^{+0.4}$& $10.0_{-0.2}^{+0.4}$ & 8.64 & $56.4_{-0.5}^{+0.3}$    & $77.6_{-0.4}^{+0.3}$ & $21.1$ &$  [16.63,19.00]        $ &$17.43^{+0.02}_{-0.03}$ & $ 	-0.32	^{+0.05	}_{-0.03} $   \\[+0.35ex]   
J124154 &0.217905  & $-$19.77 &  $138.7_{-21}^{+52}$&$11.6_{-0.2}^{+0.4}$& $10.0_{-0.2}^{+0.4}$ &8.62  & $17.4_{-1.6}^{+1.4}$    & $63.0_{-2.1}^{+1.8}$ & 94.6   &$  15.59	\pm0.12 $ &$15.72^{+0.09}_{-0.11}$ & $ 	-0.57	^{+0.16	}_{-0.09} $ \\[+0.35ex]  % ESIJune2016 8.618
J132222 &0.214431  & $-$21.18 &  $204.8_{-26}^{+44}$&$12.1_{-0.2}^{+0.3}$& $10.5_{-0.2}^{+0.3}$ & 8.80 & $57.9_{-0.2}^{+0.1}$    & $13.9_{-0.2}^{+0.2}$ &  $38.6$& $  [16.97,19.00]       $ &$19.00^{+0.01}_{-0.12}$ & $ 	-1.90	^{+0.04	}_{-0.03} $  \\[+0.35ex]  
J134251 &0.0708$^b$& $-$18.89 &  $109.1_{-16}^{+54}$&$11.4_{-0.2}^{+0.5}$& $ 9.8_{-0.2}^{+0.5}$ &8.86$^c$& $57.7_{-0.3}^{+0.3}$   & $13.9_{-0.2}^{+0.2}$ & 39.4   &$  14.61	\pm0.47 $ &$15.33^{+0.26}_{-0.69}$ & $ 	-0.02	^{+0.57	}_{-0.33} $ \\[+0.35ex]  % 8.86PPWerk  8.86 
J134251 &0.227042  & $-$21.77 &  $251.8_{-29}^{+36}$&$12.4_{-0.2}^{+0.2}$& $10.8_{-0.2}^{+0.2}$ & 8.72 & $10.1_{-10.1}^{+0.6}$    & $13.2_{-0.4}^{+0.5}$ &  $35.3$& $  18.83	\pm0.05 $ &$18.88^{+0.06}_{-0.04}$ & $ 	-0.36	^{+0.04	}_{-0.05} $ \\[+0.35ex]  
J155504 &0.189201  & $-$21.03 &  $193.7_{-25}^{+45}$&$12.1_{-0.2}^{+0.3}$& $10.5_{-0.2}^{+0.3}$ & 8.67 & $51.8_{-0.7}^{+0.7}$     & $47.0_{-0.8}^{+0.3}$ & $33.4$ & $  [16.37,19.00]      $ &$18.04^{+0.01}_{-0.90}$ & $ 	-1.43	^{+0.71	}_{-0.04} $  \\[+0.35ex]   
J213135 &0.430200  & $-$21.47 &  $199.8_{-25}^{+42}$&$12.0_{-0.2}^{+0.3}$& $10.4_{-0.2}^{+0.3}$ & 8.65 & $48.3_{-3.7}^{+3.5}$     & $14.9_{-4.9}^{+6.0}$ &  $48.4$& $  19.88	\pm0.10 $ &$19.78^{+0.01}_{-0.01}$ & $ 	-1.96	^{+0.03	}_{-0.03} $  \\[+0.35ex]  
J225357 &0.352787  & $-$20.67 &  $180.3_{-25}^{+50}$&$11.9_{-0.2}^{+0.3}$& $10.3_{-0.2}^{+0.3}$ & 8.58 & $36.7_{-4.6}^{+6.9}$     & $88.7_{-4.8}^{+4.6}$ & $203.2$& $  14.53	\pm0.05 $ &$14.56^{+0.02}_{-0.19}$ & $ <	-0.22            $ %\\[+0.35ex]    
\enddata 
\tablenotetext{a}{The full quasar name along with the quasar and galaxy RA
  and DEC can be found in \citet{pointon19}.}
 \tablenotetext{b}{Keck ESI galaxy redshifts and metallicites derived
   from this work and from
   \citet{kacprzak19,pointon19,kacprzak10}.}
 \tablenotetext{c}{Galaxy redshifts and metallicites obtained from \cite{werk12}.} 
\end{deluxetable*} 
%%%%%%%%%%%%%%%%%%%%%%%%%%%%%%%%%%%%%%%%%%%%%%%%%%%%%%%%%%%%%%%%%%

\section{SAMPLE AND DATA ANALYSIS}
\label{sec:data}

We have obtained galaxy ISM and CGM metallicities for 25 of the 47
systems selected from \citet{pointon19}, having a redshift range of
0.07$<$$z$$<$0.50 within $\sim200$~kpc (21$<$$D$$<$203~kpc) of
background quasars.  The \citet{pointon19} absorption systems were
selected based on the presence of hydrogen having a column density
range of log(N({\HI}))$=14-20$ and did not require the presence, but
must have existing spectral coverage, of metal-lines.  Our subset of
25 galaxies were selected to be star-forming such that we are able to
obtain emission-line metallicities from Keck/ESI
spectra. \citet{pointon19} selected galaxies that are isolated such
that there are no other galaxies within 100~kpc and with velocity
separations less than 500~{\kms}. From our survey, and in the
literature, the quasars fields have been surveyed to the equivalent of
a sensitivity of $\geq0.1L_*$ out to at least 350~kpc at $z=0.2$.
These {\it HST} imaged galaxy--absorber systems were identified as
part of our ``Multiphase Galaxy Halos'' Survey [from PID 13398
  \citep{kacprzak15, kacprzak19, muzahid15, muzahid16,
    nielsen17,pointon17, pointon19, ng19} and from the literature
  \citep{chen01a, chen09, prochaska11, werk12, werk13, johnson13}]. We
discuss the data and analysis below.

\subsection{Quasar Spectroscopy and Models}

The {\it HST}/COS quasar spectra have a resolution of $R\sim$20,000
and cover a range of hydrogen and metal absorption lines associated
with the targeted galaxies. Details of the {\it HST}/COS observations
are found in \citet{kacprzak15} and \citet{pointon19}.  The data were
reduced using the {\sc CALCOS} software. Individual grating
integrations were aligned and co-added using the {\sc IDL} code
`coadd\_x1d' created by
\citet{danforth10}\footnote{http://casa.colorado.edu/danforth/science/cos/costools.html}.
Since the COS FUV spectra are over-sampled, we binned the spectra by
three pixels to increase the signal-to-noise and all of our analysis
was performed on the binned spectra. Continuum normalization was
performed by fitting the absorption-free regions with smooth low-order
polynomials.

We further use Keck/HIRES or VLT/UVES quasar spectra when available to
complement our COS spectra by including coverage of {\MgI}, {\MgII},
{\FeII}, {\MnII} and {\CaII} absorption, which provide additional
metallicity constraints for absorbers with $z_{\rm abs} >$ 0.2.  HIRES
spectra were reduced using either the Mauna Kea Echelle Extraction
(MAKEE) package or IRAF. The UVES spectra were reduced using the
European Southern Observatory (ESO) pipeline \citep{dekker00} and the
UVES Post- Pipeline Echelle Reduction (UVES POPLER) software
\citep{murphy19}.

We adopted the CGM metallicities modeled from \citet{pointon19}. In
summary, the CGM metallicities were modeled using a combination of
either {\it HST}/COS or {\it HST}/COS+ Keck/HIRES or VLT/UVES spectra.
The column densities were obtained from Voigt profile fits modeled
using VPFIT \citep{carswell14}. \citet{pointon19} account for a
non-Gaussian line spread function (LSF) of the COS spectrograph by
using its wavelength dependant LSF \citep{kriss11} convolved with the
model profile during the fitting process. They assumed Gaussian LSF
for the HIRES and UVES data.  When fitting the absorption profiles,
they fit the minimum number of components to obtain a satisfactory fit
with reduced $\chi^2 \sim 1$.

The CGM metallicities are calculated in \citet{pointon19} by fitting a
grid of ionization properties generated by the ionization modeling
suite Cloudy to the calculated column densities \citep{ferland13}. We
assume a uniform single-phase layer of gas, with no dust, having solar
abundance that is irradiated by a background UV spectrum. We adopt the
HM05 UV background to generate the grids to be consistent with
previous surveys \citep{lehner13,lenher19,wotta16,wotta19}.  We used
the Markov Chain Monte Carlo (MCMC) technique described by
\citet{crighton13} to find the best-fit metallicity (quoted as the
[Si/H] ratio) and ionization parameter to the measured column
densities. The modeled N({\HI}) and CGM metallicities adopted from
\citet{pointon19} are shown in Table~\ref{tab:morph}.

\subsection{HST Imaging and Galaxy Models}

All galaxy inclination angles and galaxy-quasar azimuthal angles were
adopted from \citet{kacprzak15} and \citet{pointon19}.  All
quasar/galaxy fields have been imaged with {\it HST} using either ACS,
WFC3 or WFPC2. Details of the observations are found in
\citet{kacprzak15}. ACS and WFC3 data were reduced using the
DrizzlePac software \citep{gonzaga12} and cosmic rays were removed
during the multidrizzle process when enough frames were available,
otherwise L.A.Cosmic was used \citep{vandokkum01}. WFPC2 data were
previously reduced using the WFPC2 Associations Science Products
Pipeline (WASPP) \citep[see][]{kacprzak11b}.  Galaxy morphological
parameters were modeling with a two-component disk$+$bulge model using
GIM2D \citep{simard02}, where the disk component has an exponential
profile while the bulge has a S{\'e}rsic profile with $0.2\leq n\leq
4.0$. We apply the standard convention of an azimuthal angle
$\Phi=0^{\circ}$ defined to be along the galaxy projected major axis
and $\Phi=90^{\circ}$ defined to be along the galaxy projected minor
axis.

Galaxy photometry was adopted from \citet{kacprzak15}, who used the
Source Extractor software \citep[SExtractor;][]{bertin96} with a
detection criterion of 1.5~$\sigma$ above background.  The $m_{HST}$
magnitudes in each filter are quoted in the AB system and are listed
in Table~\ref{tab:morph}. We adopt calculated halo masses and virial
radii from \citet{ng19}, who applied halo abundance matching methods
in the Bolshoi N-body cosmological simulation \citep{klypin11} see
\citet{churchill13a,churchill13b} for further details. We then
calculate stellar masses using abundance matching models from
\citet{moster10} as described by \citet{stewart11b}.

\subsection{Galaxy Spectroscopy}

Galaxy spectra were obtained using the Keck Echelle Spectrograph and
Imager, ESI, \citep{sheinis02}.  Details of the ESI/Keck observations
are presented in \citet{kacprzak19} and \citet{pointon19}. We binned
the CCD by two in the spatial directions resulting in pixel scales of
$0.27-0.34''$ over the echelle orders of interest. Also, we binned the
CCD by two in the spectral direction resulting in a resolution of
22~\kms~pixel$^{-1}$ (${\rm FWHM}\sim90$~km/s) for a $1''$ slit.  ESI
has a wavelength coverage of 4000--10,000~{\AA}, which allows for the
detection of multiple emission lines such as {\OII} doublet,
$\rm{H}\beta$, {\OIII} doublet, $\rm{H}\alpha$, and [\NII] doublet.

All ESI data were reduced using IRAF. Galaxy spectra are both vacuum
and heliocentric velocity corrected to provide a direct comparison
with the quasar spectra.  The derived wavelength solution was verified
against a catalog of known sky-lines which resulted in a RMS
difference of $\sim0.03$~{\AA} ($\sim2$~{\kms}).  The Gaussian fitting
software \citep[FITTER: see][]{archiveI} was used to simultaneously
fit to {\Ha} and [\NII] emission lines to determine their total
flux. The line centers and velocity widths were tied together for the
two lines.  We compute a gas-phase oxygen abundance for each galaxy
using the N2 relation of \citet{pettini04}, where
12+log(O/H)=8.90+0.57$\times$N2 (N2$\equiv$log(\NII/{\Ha})). Galaxy
ISM metallicities are shown in Table~\ref{tab:morph}.
%%%%%%%%%%%%%%%%%%%%%%%%%%%%%%%%%%%%%%%%%%%%%%%%%%%%%%%%%%%%%%%%%%
\begin{figure*}
\begin{center}
\includegraphics[angle=0,scale=0.95]{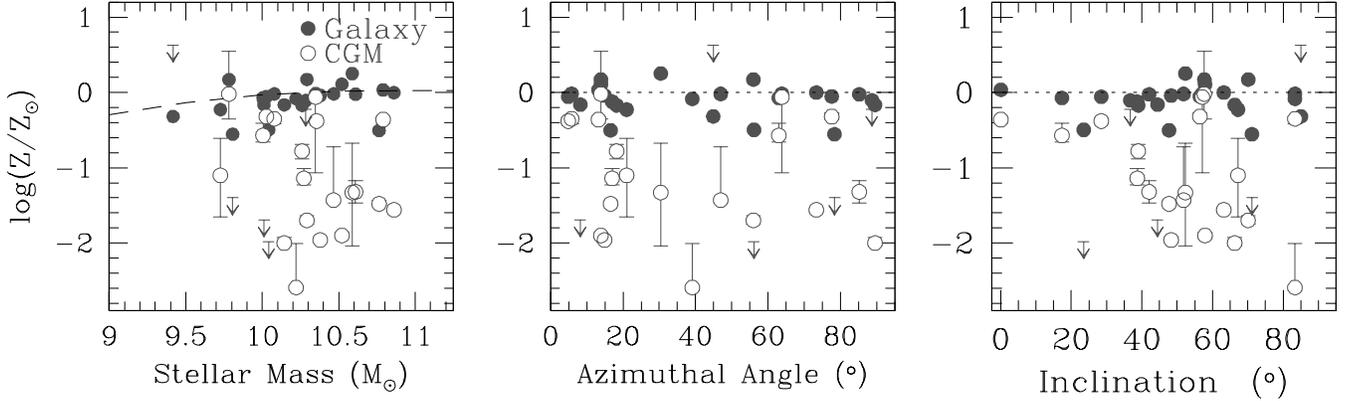}
\caption[angle=0]{(Left) The stellar mass and ISM metallicity relation
  normalized to the solar oxygen abundance of 8.69
  \citep{asplund09}. The dashed line is the expected stellar
  mass-metallicity relation at the mean redshift of our sample
  $z=0.28$ (see text for details). The CGM metallicity [Si/H] is also
  shown as a function of stellar mass, which exhibits large scatter
  relative to the ISM metallicities at fixed stellar mass.  (Middle)
  ISM and CGM metallicities as a function of azimuthal angle and
  (Right) inclination angle. As expected the ISM metallicities are
  flat as a function of azimuthal and inclination angle while the CGM
  metallicity exhibits large $\sim2$~dex of scatter.}
\label{fig:MMR}
\end{center}
\end{figure*}
%%%%%%%%%%%%%%%%%%%%%%%%%%%%%%%%%%%%%%%%%%%%%%%%%%%%%%%%%%%%%%%%%%

%%%%%%%%%%%%%%%%%%%%%%%%%%%%%%%%%%%%%%%%%%%%%%%%%%%%%%%%%%%%%%%%%%
\begin{figure}[!h]
\begin{center}
\includegraphics[angle=0,scale=1.2]{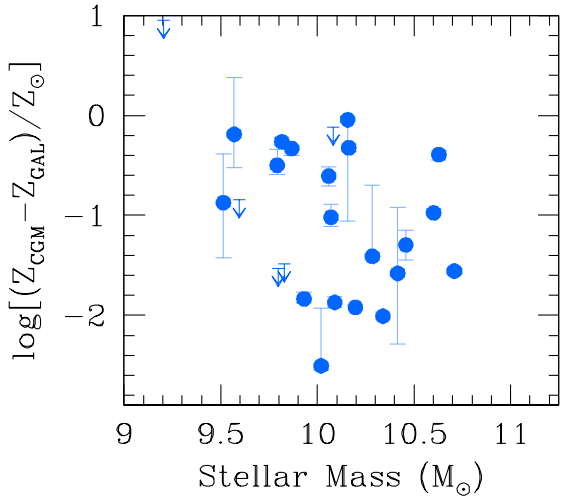}
\caption[angle=0]{The difference between the CGM and galaxy ISM
  metallicities as a function of host galaxy stellar mass.  All but
  one system has CGM metallicity higher than the galaxy
  metallicity. The 21 CGM metallicity measurements exhibit a mean
  offset from is offset from the galaxy metallicity by
  log($dZ)=-1.17\pm0.11$ where the error is quoted as the standard
  error in the mean. The scatter in this difference can be expressed
  by the standard deviation of $1\sigma=0.72$. This metallicity
  difference is independent of stellar mass over the small range
  examined here.}
\label{fig:dZvsM}
\end{center}
\end{figure}
%%%%%%%%%%%%%%%%%%%%%%%%%%%%%%%%%%%%%%%%%%%%%%%%%%%%%%%%%%%%%%%%%%
%%%%%%%%%%%%%%%%%%%%%%%%%%%%%%%%%%%%%%%%%%%%%%%%%%%%%%%%%%%%%%%%%%
\begin{figure*}
\begin{center}
\includegraphics[angle=0,scale=1.2]{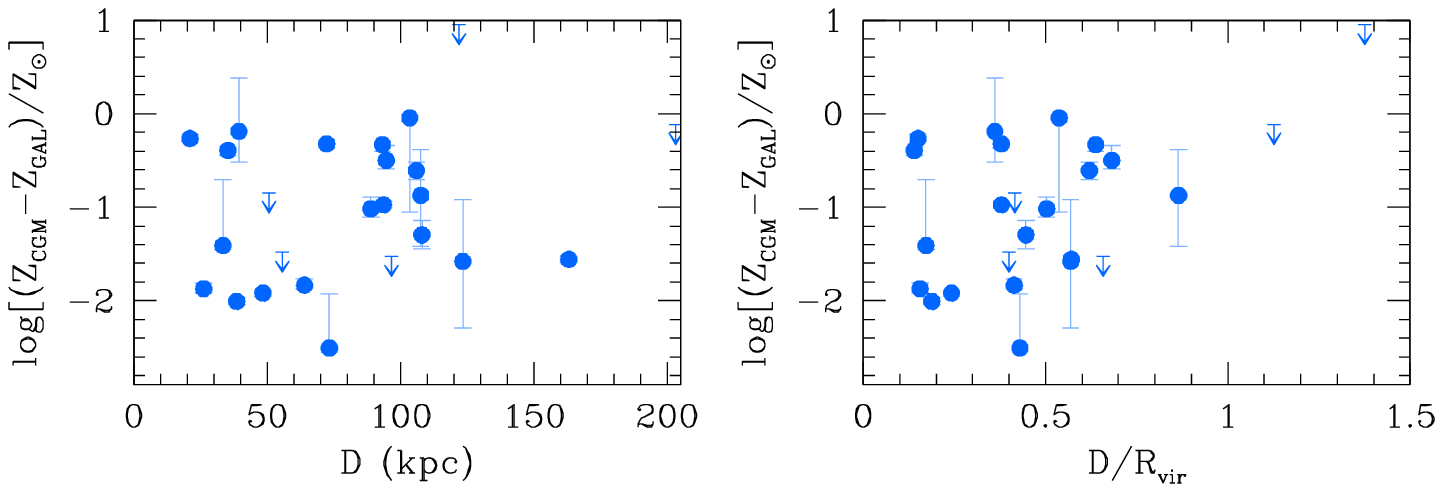}
\caption[angle=0]{The difference between the CGM and galaxy ISM
  metallicities shown as function of impact parameter (left) and as a
  function of the fraction of the virial radius (right). All of the
  CGM measurements reside within 200~kpc and most within 1~R$_{vir}$
  of their host galaxies. Arrows represent limits on the CGM
  metallicities. Note the large scatter at all distances away from the
  galaxy with no obvious metallicity gradient.}
\label{fig:ZvsD}
\end{center}
\end{figure*}
%%%%%%%%%%%%%%%%%%%%%%%%%%%%%%%%%%%%%%%%%%%%%%%%%%%%%%%%%%%%%%%%%%

\section{Results}\label{sec:results}

In this section we explore the metallicities of both the galaxy ISM
and of the CGM to determine if there is a relationship between distant
CGM gas and its host galaxies.

In Figure~\ref{fig:MMR}, we present galaxy ISM metallicities as
determined from the {\Ha} and {\NII} line ratios normalized to oxygen
solar abundance of 8.69 \citep{asplund09}. Here log($Z$) is defined
for galaxies as the ratio of the mass of oxygen in the gas-phase and
the hydrogen gas mass. Our sample of galaxies have a stellar mass
range from 9.4$\leq$log(M$_*$/M$_{\odot}$)$\leq$10.9 with roughly
0.5~dex error on the stellar mass. The dashed line shows the galaxy
mass-metallicity relation obtained from the formalism of
\citet{zahid14} evaluated at our mean galaxy redshift of $z=0.28$ and
then normalized to the solar oxygen abundance. We also normalize the
\citet{zahid14} relation to the N2 \citet{pettini04} calibration used
to calculate our galaxy ISM metallicities following the methods of
\citet{kewley08}\footnote{ Note \citet{zahid14} used the
  \citet{kobulnicky04} calibration and the difference between these
  calibration methods can lead to offset of $\sim$0.3~dex in
  metallicity.}.

We find that our galaxy ISM metallicities agree with the expectations
and follow the general trend of increasing metallicity with increasing
mass having a $1\sigma$ scatter of 0.19~dex about the relation. This
scatter could be further reduced if the mass-metallicity relation was
computed for the full range of galaxy redshifts observed here, however
this is beyond the scope of this paper and not necessary for our
analysis.

While the galaxy metallicities exhibit a tight relation with stellar
mass (0.19~dex scatter), the metallicity of the CGM shows no
dependence with stellar mass and exhibits a scatter that ranges over
2~dex. This clearly shows that the CGM is more complex and the
metallicity is likely driven by a range of processes compared to that
of the ISM. In all cases, except for a poorly constrained limit, the
CGM metallicity is always lower than the galaxy ISM
metallicity. Figure~\ref{fig:dZvsM} shows the difference between the
CGM and galaxy ISM metallicities as a function of host galaxy stellar
mass. Using a survial analysis with all of the data we find that the
CGM metallicity is offset from the galaxy metallicity by
log($dZ)=-1.17\pm0.11$ where log($dZ$) is quoted as the mean offset
from the galaxy metallicity while the error is quoted as the standard
error in the mean. The scatter in this difference can be expressed by
the standard deviation of $1\sigma=0.72$. This metallicity difference
is independent of stellar mass over the small range examined here.
The relative CGM and ISM metallicities for the stellar mass range of
9.7$\leq$M$_*<$10.3 and 10.3$\leq$M$_*\leq$10.8 exhibit metallicity
differences of $-1.27\pm0.14$ (1$\sigma$ scatter of 0.91) and
$-1.09\pm0.17$ (1$\sigma$ scatter of 0.67), respectively, with values
quoted as the mean difference while the error is quoted as the
standard error in the mean. We have also applied generalized Kendall
and Spearman rank correlation tests, which accounts for measured
limits in the sample \citep{feigelson85}, between the stellar mass and
log($dZ$).  We no strong supporting evidence for trends between
stellar mass and log($dZ$) ($2.1\sigma$ -- Kendall, 2.3$\sigma$ --
Spearman).

Given that our sample is low redshift ($\left<z\right>=0.28$), where
one could expect metal-poor accretion to minimal and metals within the
CGM could be well mixed or metal enriched from Gyrs of ongoing
outflows, there is still a significant metallicity difference between
the host galaxy and the CGM. Furthermore, this difference is
independent of stellar mass.

The middle and right panels of Figure~\ref{fig:MMR} show both the
galaxy and CGM metallicities as a function of the azimuthal and
inclination angles, respectively. As expected, the galaxy ISM
metallicity is independent of the galaxy orientation with respect to
the quasar sight-line as well as the galaxy's inclination angle. The
CGM however exhibits large scatter as a function of azimuthal angle
and inclination angle as previously shown by \citet{pointon19} using a
larger sample of 47 galaxy-absorber pairs. \citet{pointon19} explored
how the CGM metallicity behaves relative to the galaxy inclination and
azimuthal angles and found no apparent trend, which conflicts with a
scenario of planer accretion and bi-polar outflows
\citep[e.g.,][]{nelson19}. However, the Pointon et al.\ study did not
address how the relative galaxy-CGM metallicity behaves as a function
of orientation or impact parameter.

%%%%%%%%%%%%%%%%%%%%%%%%%%%%%%%%%%%%%%%%%%%%%%%%%%%%%%%%%%%%%%%%%%
\begin{figure*}
\begin{center}
\includegraphics[angle=0,scale=1.2]{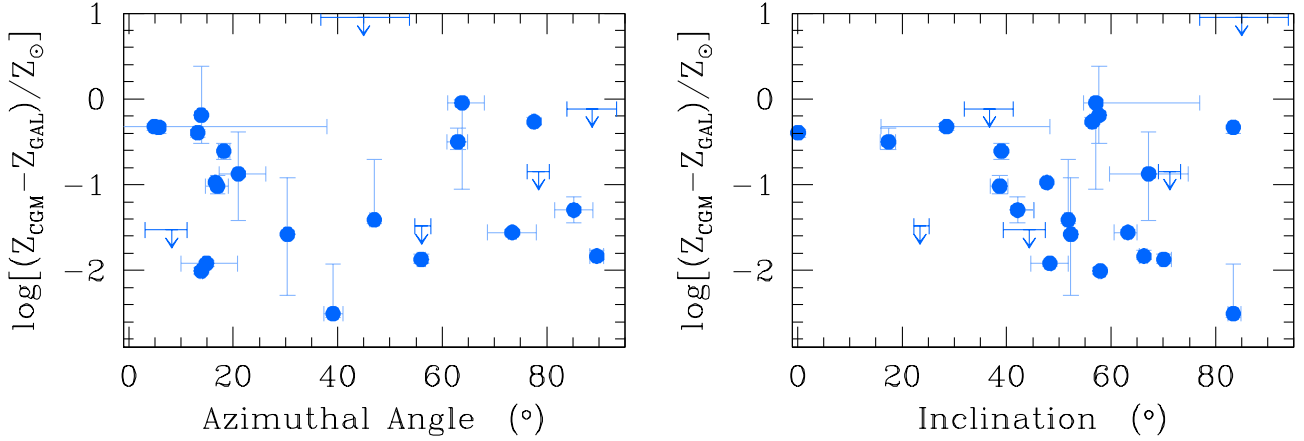}
\caption[angle=0]{The difference between the CGM and galaxy ISM
  metallicities as a function of azimuthal angle (left) and galaxy
  inclination (right). In a simple CGM model, we would expect low
  metallicity gas to accrete along the major axis (bottom left corner
  of the plot) while higher metallicity gas outflows along the minor
  axis (upper right corner of the plot). When taking the galaxy
  metallicity into account, we do not find a correlation with
  azimuthal angle as expected from the simple model. The exists a
  range of metallicities at all azimuthal angles. (Right) As a galaxy
  becomes more edge-on, it is expected that outflows and accretion
  signatures would be more apparent than for near face-on galaxies. We
  do find that highly inclined galaxies have a range of metallicities
  that could arise from outflow and/or accretion signatures.}
\label{fig:ZvsOri}
\end{center}
\end{figure*}
%%%%%%%%%%%%%%%%%%%%%%%%%%%%%%%%%%%%%%%%%%%%%%%%%%%%%%%%%%%%%%%%%%

It is unclear how we expect the CGM metallicity to behave as a
function of impact parameter. Simulations predict co-planer accretion
with bi-conical outflows \citep[e.g.,][]{nelson19} and negative radial
metallicity gradients for both outflow and accretion models
\citep[e.g.,][]{freeke12}.  Furthermore, simulations of extended disk
ISM metallicities show that galaxies have negative metallicity
gradients extending out to 10--20~kpc
\citep[e.g.,][]{kobayashi11,pilkington12}, while observations show
either flat or negative gradients with significant scatter in their
slopes \citep[e.g.,][]{wuyts16,sanchez18}.

%%%%%%%%%%%%%%%%%%%%%%%%%%%%%%%%%%%%%%%%%%%%%%%%%%%%%%%%%%%%%%%%%%
\begin{figure*}
\begin{center}
\includegraphics[angle=0,scale=1.2]{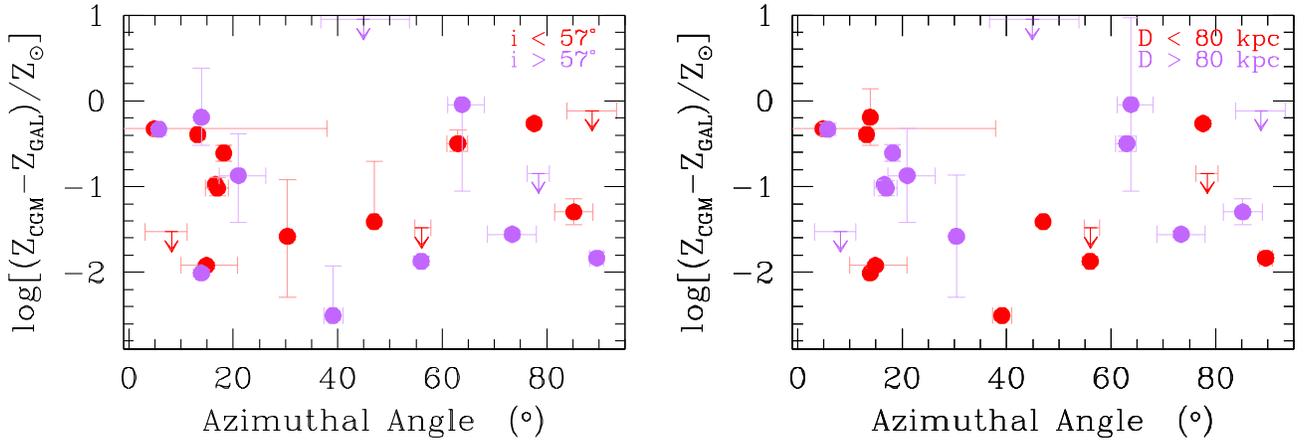}
\caption[angle=0]{Same as left panel of Figure~\ref{fig:ZvsOri} except
  now the data are color-coded as a function high and low inclination
  angles (left) and high and low impact parameters (right). Note that
  regardless of low or high inclination, or low and high impact
  parameter, there is no correlation with the relative ISM-CGM
  metallicity and azimuthal angle.}
\label{fig:ZvsOriC}
\end{center}
\end{figure*}
%%%%%%%%%%%%%%%%%%%%%%%%%%%%%%%%%%%%%%%%%%%%%%%%%%%%%%%%%%%%%%%%%%
%%%%%%%%%%%%%%%%%%%%%%%%%%%%%%%%%%%%%%%%%%%%%%%%%%%%%%%%%%%%%%%%%%
\begin{figure}
\begin{center}
\includegraphics[angle=0,scale=1.2]{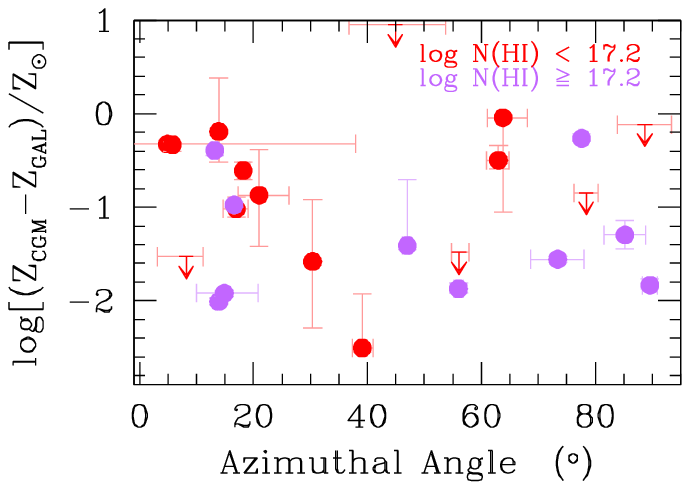}
\caption[angle=0]{Same as Figure~\ref{fig:ZvsOriC} except now the data
  are color-coded as a function of low (log N({\HI})<17.2) and high
  (log N({\HI})$\geq$17.2) modeled {\HI} column densities.}
\label{fig:ZvsOriC_NH}
\end{center}
\end{figure}
%%%%%%%%%%%%%%%%%%%%%%%%%%%%%%%%%%%%%%%%%%%%%%%%%%%%%%%%%%%%%%%%%%

In the left panel of Figure~\ref{fig:ZvsD}, we present the difference
in metallicity between the CGM and galaxy as a function of impact
parameter. All of our absorption systems are within 200~kpc of the
host galaxy.  We find no apparent metallicity gradient ($0.78\sigma$ --
Kendall, 0.92$\sigma$ -- Spearman) with a larger scatter of low and
high metallicity systems at all distances away from the galaxy.  It is
interesting to note that the two lowest metallicity systems reside
within 75~kpc (or within 0.5$R_{vir}$) of the host galaxy, which is
counter-intuitive since one might expect these more unpolluted systems
to reside further away from their host galaxies (unless metal-poor
accretion is really reaching low impact parameters without mixing).

It is more meaningful to show the difference in ISM/CGM metallicities
as a function of the the fraction of the virial radius given that
these galaxies cover a range of halo masses. In the right panel of
Figure~\ref{fig:ZvsD} shows the difference between CGM and ISM
metallicities as a function of the fraction of the virial
radius. Almost all of our absorption systems, except for two limits,
reside within 1~R$_{vir}$ of their host galaxies. Again, we find no
strong trend between metallicity and the fraction of the virial radius
($1.63\sigma$ -- Kendall, 1.51$\sigma$ -- Spearman). While the most
metal poor absorbers reside near to the galaxy, there is significant
scatter at all radii.  The scatter seen here could be a result of gas
being enriched from outflows, while metal-poor gas could come from
accreting gas. All of these might be expected to have an orientation
dependence.

In the left panel of Figure~\ref{fig:ZvsOri} we present the difference
between the CGM and galaxy ISM metallicities as a function of
azimuthal angle. In a simple CGM scenario, one may expect that
metal-poor gas relative to the galaxy should accrete along the major
axis of the galaxy disk, which should populate the lower left corner
of the plot. On the other hand, metal-enriched outflows relative to
the host galaxy should occur along the galaxy minor axis, which should
populate the upper right corner of the plot. However, it is clear from
the figure that there is a large range in log($dZ$) from $0$ to $-$2
at all azimuthal angles. We find that the mean metallicity difference
between the CGM and host galaxy for both along the major and minor
axes, bifurcated at 45~degrees, are $-$1.13$\pm$0.18 (1$\sigma$ scatter
of 0.76) and $-$1.23$\pm$0.11 (1$\sigma$ scatter of 0.65),
respectively.

It does seems clear that there is no significant relative metallicity
dependence as a function of azimuthal angle. This is consistent with
the results of \citet{pointon19} who showed that the CGM metallicity
alone does not have a metallicity dependence. When taking into account
the host galaxy metallicity, this does not unveil a new result.  This
is consistent with the first suggestive results of \citet{peroux16}
using 9 galaxy absorber pairs. Given that there exists a large scatter
in CGM metallicities, while ISM metallicities show very little
scatter, then it is not surprising no additional relationships are
discovered here. It is also interesting to note that the most metal
enriched systems are along the major axis of the galaxy and not the
minor axis, where outflows are expected to dominate. So it is unclear
what is the source of these high metallicity systems and/or if they
are part of a very extended HI disk. It is plausible that recycled
metal-enriched gas could fall back towards the galaxy and reaccrete
along the major axis, which could explain the large scatter seen at
low azimuthal angles. However, it is puzzling that there exists
absorption systems that have much lower metallicities than their host
galaxies along the minor axis.

The right hand panel of Figure~\ref{fig:ZvsOri} shows the relative
metallicity as a function of galaxy inclination. We do not have many
near face-on galaxies in our sample, so we are unable to comment on
the distribution of metallicities here. However, in a simple
inflow/outflow scenario, one would assume that these gas flows may be
more distinguishable for edge-on galaxies. At intermediate to highly
inclined galaxies, we find significant scatter ranging from 0$\lesssim
$log (dZ)$ \lesssim -2$. This further indicates that there is likely
no metallicity dependence on galaxy inclination.

It could be possible that combination of inclination angles and/or
impact parameters could dilute any correlation between the relative
galaxy ISM an CGM metallicities as a function of azimuthal angle. In
Figure~\ref{fig:ZvsOriC} we explore the azimuthal dependence of the
relative metallicity bifurcated by high and low galaxy inclination
angles at 57 degrees, which splits the sample roughly equally into two
subsets. For highly inclined galaxies, we would expect to see the
strongest relation between relative metallicity and azimuthal angle
since the quasar sight-line should only pass though individual outflow
and accretion structures and not a blend of the two in projection. We
find a similar scatter in metallicity for both low and high
inclination galaxies. Interestingly, we find the lowest metallicities
relative to their host galaxies tend to be highly inclined and exist
over the full range of azimuthal angles. Low inclination galaxies have
fewer very metal-poor CGM systems which could be due to an averaging
of structures and gas metallicities along the quasar sight-line (i.e,
passing though both outflows and accreting gas).

The right panel of Figure~\ref{fig:ZvsOriC} also shows the azimuthal
metallicity dependence as a function of low ($D<80$~kpc) and high
($D>80$~kpc) impact parameters. Since it has been shown that outflows
may only extend out to 50--100~kpc \citep[e.g.,][]{bordoloi11,lan18},
one could expect the highest metallicity systems, or at least
metal-enriched systems (at or above the galaxy ISM metallicity), to
exist at low impact parameters and possibly along the galaxy minor
axis. We find that along the minor axis, both low and high impact
parameter systems have a range of metallicities. In fact, the three
low metallicity systems are at low impact parameters, which is
unexpected. Again, along the major axis, both low and high impact
parameter systems have a range of relative CGM to galaxy
metallicity. Therefore impact parameter doesn't seem to play a
critical role in the azimuthal dependence for the metallicity
difference between the CGM and the host galaxies. We do find that
intermediate azimuthal angles are dominated by $D<80$~kpc systems,
where possibly extended disk or interactions may also contribute to
the absorption detected here.

It is further possible that any azimuthal dependence could be driven
by the hydrogen column density since the CGM metallicity bi-modality
is only shown for pLLSs and LLSs \citep{wotta16,wotta19}. In
Figure~\ref{fig:ZvsOriC_NH} we show the ISM-CGM metallicity difference
versus azimuthal angle separated into high N({\HI})
(logN({\HI})$\geq17.2$ -- purple) and low N({\HI}) (logN({\HI})$<17.2$
-- red).  In this figure high column density systems tend to have
lower metallicities, which is consistent with previous work
\citep[see][]{pointon19}. The vast majority of low column density
systems (10/15 -- red points in Figure~\ref{fig:ZvsOriC_NH}) reside
along the galaxy major axis and only four low column density systems
with metal-line measurements are found at greater than
$\sim60$~degrees. This could be suggestive that low N({\HI}) systems
are better tracers of accretion if the accreting gas has a range of
metallicities, however more data is required.  It is also possible
that low N({\HI}) gas along the major axis of galaxies occurs as both
metal-poor gas accretion and metal-enriched recycling.

We also find that 6/10 high column density systems (purple points in
Figure~\ref{fig:ZvsOriC_NH}) reside above an azimuthal angle of
40~degrees, suggesting that high column density systems could better
trace outflows. However, the metallicities along the major and minor
axes are consistent with each other.

\section{Discussion}\label{sec:discussion}

Simulations clearly show that the CGM is complex, yet observations
have shown that the spatial distribution of high and low ions are
azimuthally dependent
\citep{bouche12,kacprzak12a,lan14,kacprzak15,lan18}. Even the internal
dispersion of the CGM absorption for low ions points to accretion and
outflow scenarios \citep{nielsen15}. Furthermore, relative gas and
galaxy kinematics show that low ions are kinematically connected to
their host galaxy by aligning with their rotation curves and being
modelled well by accreting+corotating gas
\citep{steidel02,kacprzak10,kacprzak11a,ho17,martin19,zabl19}. On the other
hand, minor axis gas also seems to be well modelled by outflowing gas
\citep{bouche12,gauthier12,schroetter16}. Finally, the metallicity
distribution of LLS and pLLS appears bimodal, which also suggests that
outflows and accretion are dominant phenomena within the CGM
\citep{wotta16,wotta19}.  Thus the spatial distribution of metallicity
around galaxies seemed to be key to understanding the origins of the
CGM.

However, as \citet{pointon19} has shown, CGM metallicity alone has no
correlation with azimuthal angle or inclination regardless of impact
parameter, N({\HI}), etc. This is quite disappointing given the simple
picture presented by observations. However, we do not know how the
relative galaxy-ISM and CGM metallicities affect these results given
the well-known galaxy stellar mass and ISM metallicity relation found
at all redshifts
\citep[e.g,][]{tremonti04,sanders14,steidel14,zahid14,kacprzak15b}. Thus
accounting for the galaxy metallicity could enhance any possible
relationship with metallicity and galaxy orientation. Here we examine
these relationships using 25 systems with both galaxy ISM and CGM
metallicities.

We find that although host galaxies follow a stellar mass metallicity
relation (0.19~dex scatter over the mass range
9.4$\leq$log(M$_*$/M$_{\odot}$)$\leq$10.9), the CGM is quite scattered
as a function of stellar mass spanning 2~dex in metallicity.  This is
expected as galaxy ISM metallicities are driven by stellar evolution
and gas accretion and are averaged over entire galaxy disks while the
CGM detected along point-like quasar sightlines may originate from IGM
gas accretion, from nearby galaxies/satellites, or from recycled and
outflowing gas generated from within the galaxy.  We find that the
mean of the CGM metallcities are lower than the mean galaxy
metallicities by $-1.17\pm0.11$. This offset is independent of stellar
mass over the small range examined here.  The mean CGM metallicities
for stellar mass ranges of9.7$\leq$log(M$_*/$M$_{\odot}$)$_*<$10.3 and
10.3$\leq$ log(M$_*/$M$_{\odot}$)$_*\leq$10.8 are lower than the
galaxy metallicity by $-1.27\pm0.14$ (1$\sigma$ scatter of 0.91) and
$-1.09\pm0.17$ (1$\sigma$ scatter of 0.67), respectively.  Thus there
is a significant difference between the host galaxy and the CGM
metallicities, which are stellar mass independent.  There may be a
small hint of a correlation with stellar mass and log($dZ$) but it is
not highly significant ($2.1\sigma$ -- Kendall, 2.3$\sigma$ --
Spearman).

These results are consistent with the findings of \citet{prochaska17}
who has shown that the CGM metallicity does not correlate with the ISM
metallicity of host galaxies nor does the CGM metallicity correlate
with stellar mass. However, it is difficult to compare our works
directly since they use the \citet{haardt12} (HM12) ionizing
background. Previous works have shown that harder spectrum of ionizing
photons from the HM12 background is due to a lower escape fraction of
radiation from galaxies compared to the HM05 background, which leads
to higher metallicity estimates and an anti-correlation between
N({\HI}) and metallicity \citep{howk09,werk14,wotta16,wotta19,chen17,
  zahedy19, pointon19}.

All of our CGM metallicities are lower than the galaxy ISM
metallicities. This could imply that CGM may originate from a nearby
satellite galaxy or its outflow or tidal debris. However, the halo gas
cross-section of satellite galaxies are predicted to be extremely
small \citep{gauthier10,martin12,tumlinson13} and thus, an unlikely
contributor to the bulk of the detected absorption.

The lower CGM metallicities could imply that gas ejected from galaxies
is diluted with metal poor gas within the CGM or metals ejected from
host galaxies have taken a long time to travel out into the CGM. If
the gas does take a long time to travel out into the CGM, an
interesting experiment is to then see at what age of the Universe did
a $z=0.28$ M$_{*}\sim$10$^{10.5}$M$_{\odot}$ galaxy have a ISM
metallicity that was $-$1.2~dex lower than its current value.  We
estimate this gas must have been ejected prior to $z=3$ given the
limits of the stellar mass evolution of $\sim$10$^{10.5}$M$_{\odot}$
galaxy \citep{papovich15} and combined with the evolution of the
mass-metallicity relation \citep{mannucci09,zahid14}. Thus, the gas
would have to be ejected roughly at $>$8~Gyr prior to $z=0.28$ in
order to have a galaxy such a low metallicity. This large time-scale
provides ample time for ejected gas to travel out into the CGM and
also recycle back to the disk since this is estimated to take at least
1 Gyr \citep[e.g.,][]{oppenheimer09,oppenheimer10}. However, this also
assumes no gas mixing, which would likely further change the
metallicity of the ejected material.  So it seems possible that any
metal poor gas that was ejected at early times should have been
enriched several times over a $>$8~Gyr time-frame.  Thus, in order to
find low metallicity systems along the minor axis of galaxies,
outflowing gas would have to be well-mixed with its metal-poor
surroundings within the CGM or maybe the cool CGM is not a good tracer
of galactic outflows.

\citet{peroux16} first looked into the difference between the galaxy
ISM and CGM metallicities as a function of azimuthal angle with nine
galaxies and suggested that there seems to be a large scatter along
the major axis while they only had two lower relative metallicity
systems along the minor axis.  We further find that the mean
metallicity differences along the major and minor axes, bifurcated at
45~degrees, are $-$1.13$\pm$0.18 (1$\sigma$ scatter of 0.76) and
$-$1.23$\pm$0.11 (1$\sigma$ scatter of 0.65), respectively.
Regardless of whether we examine our sample by low/high inclination or
low/high impact parameter, or low/high column density (or any
combination of these), we do not find any significant relationship
with relative metallicity and azimuthal angle.

So what is going on here and should we be focusing on metallicity when
examining modes of accretion and outflows? Outflows do occur and there
is plenty of evidence that they likely occur along the minor axis and
this gas has to be metal-enriched. Also, some form of accretion must
happen given all the kinematic evidence found for CGM-galaxy pairs and
that fact that galaxies continue to form stars. Yet it is unclear what
the metallicity of that accreting gas could be. Cosmological
simulations predict that gas accretion metallicities range between
10$^{-3}-10^{-0.5}$~~$Z_\odot$, which is dependent on redshift and
halo mass
\citep{keres05,fumagalli11b,oppenheimer12,freeke12,shen13,kacprzak16},
however this range does overlap with the expected metallicities of
recycled/outflowing gas.  Also, the complexity of outflowing gas makes
things worse given there is typically hot outflowing material
containing cool entrained clouds. Thus maybe metallicity alone is a
poor indicator of the origins of the CGM gas or the metallicity of low
ions might be a poor indicator of the metallicity of hot outflowing
gas.

Analysis of cosmological simulations from \citet{ford14} showed that
low-ionization metal absorbers tend to arise within inflowing gas,
while high-ionization metal absorbers trace ancient outflowing gas
deposited in galaxy haloes many Gyr ago. \citep{muzahid15} showed a
galaxy having both a metal-poor low ionization component ($\sim$-1.5)
and a high ionization metal rich component ($>0.3$). They concluded
that the low ionization metal-poor phase was consistent with being
recycled material in the galaxy halo and that the high-ionization,
metal-enriched, low density gas presumably originated from
star-formation driven outflows from the host-galaxy. Thus it is
possible that different gas phases have different origins and given
this example, more work is needed to further model the multi-phase CGM
metallicities.

It is still puzzling however that pLLSs and LLSs have bimodal
metallicity distribution and this needs to be explored within
simulations and determined whether the bimodality is caused by
internal CGM properties or due to environmental effects. So far
cosmological simulations have been unable to reproduce the CGM
metallicity bimodality
\citep{hafen17,hafen18,rahmati18,lehner19}. Furthermore environmental
effects may not be the likely mechanism producing the bimodality
either \citep{pointon19b}. It seems that properties such as
velocities, column densities and equivalent widths that are straight
forward to measure provide the most fruitful evidence for gas flows.

On the other hand, modelling the total metallicity along a given
sight-line is not straightforward either and can lead to confusing
results. We know that the CGM metallicity must vary along the
sight-lines \citep{churchill15,peeples18}, however most studies model
a global single-phase metallicity since it is difficult in most cases
to assign the correct amount of hydrogen to given metal features from
different gas phases in a single spectrum.  \citet{lehner19} has shown
for $\sim30$ near redshift-separated absorbers (separations of
50--400~\kms) have metallicities differences ranging from 0 to 1.7dex.
Only a smaller number of absorption-line systems have been modeled as
multi-phase and with cloud-to-cloud metallicities
\citep[e.g.,][]{prochter10,tripp11,crighton13b,muzahid15,muzahid16,rosenwasser18,zahedy19}. Furthermore,
it is expected that absorption arising from outflows would have large
cloud-to-cloud variations in metallicity and ionization level
\citep[e.g.,][]{veilleux05,rosenwasser18,zahedy19}, which we are
averaging over. We are also metal-biased in that detecting some metals
at a given velocity does not imply there is no metal poor gas at that
same velocity in some other spatial location along the sight-line that
is masked by those other metal lines. Maybe the CGM is not well mixed
and metallicity is not a great indicator of dynamic processes and we
are best to focus our efforts on dynamic/kinematic measurements to
study gas flows.

Either way, larger and well targeted samples may provide future
insight to the metallicity distribution around galaxies.

\section{Conclusions}\label{sec:conclusion}

We present galaxy ISM and CGM metallicities for 25 absorption systems
associated with isolated star-forming galaxies ($0.07\leq
z\leq0.50$). Galaxy ISM metallicities were measured using {\Ha} and
[\NII] emission lines obtained from Keck/ESI spectra. The CGM
metallicities were adopted from \citet{pointon19}, which were modeled
using an MCMC analysis along with Cloudy. We examine the galaxy mass
metallicity relation for our galaxies and their absorption systems. We
also explore whether the relative galaxy ISM and CGM metallicity
correlates with galaxy orientation with respect to the quasar. Our
results are summarized as follows:

\begin{enumerate}

\item We find that our galaxy ISM metallicities agree with the
  expectations of following the general trend of increasing
  metallicity with increasing stellar mass having a $1\sigma$ scatter
  of 0.19~dex about the relation determined at
  $\left<z\right>=0.28$. This scatter could be further reduced if the
  mass-metallicity relation was computed for the full range of galaxy
  redshifts in our sample.

\item CGM metallicity shows no dependence with stellar mass ($<2.3
  \sigma$ significance) and exhibits a scatter that ranges over 2~dex.
  The CGM and galaxy galaxy metallicity differences for stellar mass
  ranges of 9.7$\leq$M$_*<$10.3 and 10.3$\leq$M$_*\leq$10.8 are
  $-1.27\pm0.14$ (1$\sigma$ scatter of 0.91) and $-1.09\pm0.17$
  (1$\sigma$ scatter of 0.67), respectively.  Thus, even at low
  redshift, where one might expect the global metallicities to be more
  homogenized, there is still a significant difference between the
  host galaxy ISM and the CGM metallicities and are stellar mass
  independent

\item The CGM metallicities are always lower than
  the galaxy ISM metallicities and are offset by log($dZ)=-1.17\pm0.11$
  where log($dZ$) is quoted as the mean offset from the galaxy metallicity
  while the error is quoted as the standard error in the mean. The
  scatter in this offset can be expressed by the standard deviation of
  $1\sigma=0.72$.

\item All of our CGM measurements reside within 200~kpc and
  1.5~R$_{vir}$ of their host galaxies. We find no obvious metallicity
  gradient as a function of impact parameter or virial radius ($<1.6
  \sigma$ significance). This could be diluted with a range of galaxy
  orientations within that sample. Ideally, this sort of work would
  best be done for a large sample of edge-on galaxies.

\item There is no relative CGM--galaxy metallicity as a function of
  azimuthal angle. We find that the mean metallicity differences along
  the major and minor axes, bifurcated at 45~degrees, are
  $-$1.13$\pm$0.18 (1$\sigma$ scatter of 0.76) and $-$1.23$\pm$0.11
  (1$\sigma$ scatter of 0.65), respectively.

\item Regardless of whether we examine our sample by low/high
  inclination or low/high impact parameter, or low/high {\HI} column
  density (or any combination of these), we do not find any
  significant relationship with relative CGM--galaxy metallicity and
  azimuthal angle.

\item 
The majority of low column density systems (10/15 --
logN({\HI})$<17.2$) reside along the galaxy major axis and only two
low column density systems with metal-line measurements are found at
$\sim60$~degrees.  We also find that 6/10 high column density systems
(logN({\HI})$\geq 17.2$) reside above an azimuthal angle of
40~degrees, suggesting that high column density systems could better
trace outflows. However, the metallicities along the major and minor
axes are consistent. This could be suggestive that low N({\HI})
systems are better tracers of accretion if the accreting gas has a
range of metallicities. More data is required to determine whether
these trends really do exist.

\end{enumerate}

It is undoubtedly true that the CGM is complex. The community has put
forth a large body of work showing evidence for accretion and
outflows, however a clear confirmation of cosmological accretion
remains elusive. CGM metallicities and metallicity differences between
the galaxy-ISM to CGM do not help illuminate our understanding of the
CGM, at least with current sample sizes. We further need to address how
assuming averaged line of sight metallicities and/or single phase
metallicities truly effects our results.

An additional issue is that our point-source quasars probe through
individual galaxy halos, which could give rise to large variations in
metallicity within the halo and along the sight-line. Hopefully in the
future, we will be able to use background galaxies, or gravitational
lenses \citep[e.g.,][]{lopez18} to obtain a better sampling of the
halo metallicities and to be less susceptible to line-of-sight
variations. For now, it seems that properties such as velocities,
column densities and equivalent widths that are easy to measure
provide the most fruitful evidence for gas flows.\\

\acknowledgments We thank Roberto Avila (STScI) for his help and
advice with modeling PSFs with ACS and WFC3. GGK and NMN acknowledges
the support of the Australian Research Council through a Discovery
Project DP170103470. Parts of this research were supported by the
Australian Research Council Centre of Excellence for All Sky
Astrophysics in 3 Dimensions (ASTRO 3D), through project number
CE170100012. CWC and JCC are supported by NASA through grants HST
GO-13398 from the Space Telescope Science Institute, which is operated
by the Association of Universities for Research in Astronomy, Inc.,
under NASA contract NAS5-26555. CWC and JCC are further supported by
NSF AST-1517816.  Most of the data presented here were obtained at the
W. M. Keck Observatory, which is operated as a scientific partnership
among the California Institute of Technology, the University of
California, and the National Aeronautics and Space Administration. The
Observatory was made possible by the generous financial support of the
W. M. Keck Foundation.  Observations were supported by Swinburne Keck
programs 2016A\_W056E, 2015B\_W018E, 2014A\_W178E and
2014B\_W018E. The authors wish to recognize and acknowledge the very
significant cultural role and reverence that the summit of Mauna Kea
has always had within the indigenous Hawaiian community.  We are most
fortunate to have the opportunity to conduct observations from this
mountain. Based on observations made with the NASA/ESA Hubble Space
Telescope, and obtained from the Hubble Legacy Archive, which is a
collaboration between the Space Telescope Science Institute
(STScI/NASA), the Space Telescope European Coordinating Facility
(ST-ECF/ESA) and the Canadian Astronomy Data Centre (CADC/NRC/CSA).

%% Included in this acknowledgments section are examples of the
%% AASTeX hypertext markup commands. Use \url without the optional [HREF]
%% argument when you want to print the url directly in the text. Otherwise,
%% use either \url or \anchor, with the HREF as the first argument and the
%% text to be printed in the second.

%doing the math in section~\ref{bozomath}.
%More information on the AASTeX macros package is available \\ at
%\url{http://www.aas.org/publications/aastex}.
%For technical support, please write to
%\email{aastex-help@aas.org}.


%% in parentheses, after the keyword, but they will not be verified.

{\it Facilities:} \facility{Keck II (ESI)}
\facility{HST (COS, WFPC2, ACS, WFC3)}.

\end{document}